\documentclass{openjournal}
\usepackage{graphicx}

\usepackage[plainpages=false, colorlinks=true, anchorcolor=blue, linkcolor=blue, citecolor=blue, bookmarks=false]{hyperref}
\usepackage{color}
\usepackage{float}
\usepackage{amsfonts,amsmath,amssymb}
\usepackage{array}
\usepackage{subfigure}
\usepackage{subcaption}
\usepackage{caption}
\usepackage{booktabs}
\usepackage{aas_macros}
\usepackage{multirow}
\newcommand{\rthis}[1]{\textcolor{black}{#1}}
\begin{document}
\title{ A study of gamma-ray emission from OJ 287 using Fermi-LAT from 2015-2023 }
\author{Vibhavasu Pasumarti}
\email{ep20btech11015@iith.ac.in}
\author{Shantanu Desai}
\email{shntn05@gmail.com}
\affiliation{Department of Physics, IIT Hyderabad, Kandi Telangana-502284, India}

\begin{abstract}
We search for gamma-ray emission from OJ287 in the energy range from 0.1-300 GeV during 2015-2023, in coincidence with an extensive observing campaign to monitor the optical flux variability and polarization, as discussed in arXiv:2311.02372. We present results for eight segments in the aforementioned period, with each segment corresponding to an observing season. We report non-zero gamma-ray flux (with $>10\sigma$ significance) for all the eight segments. The photon and energy flux observed during this period is $\sim 5.2 \times 10^{-8} \rm{ph~cm^{-2}~s^{-1}}$ and $2.75 \times 10^{-5} \rm{MeV~cm^{-2}~s^{-1}}$, respectively, while the SED can be fitted with a power law with a slope of $2.12 \pm 0.02$. The observed luminosity is between $10^{45}-10^{46}$ ergs/sec. Similar to previous works we find a hard cutoff in the gamma-ray spectrum at around 20 GeV. We do not observe any correlations between the peaks in the gamma ray light curves and corresponding optical light curves in $I$ and $R$ bands, as well as the  X-ray light curves from SWIFT between 0.3-10 keV. 

 \end{abstract}
\maketitle
\section{Introduction}
OJ287 is a BL object located at a redshift of z=0.306 ($D_L \sim 1600$~Mpc) ~\citep{Sitko}. This object is known to contain a supermassive black hole binary and hence has been a promising target for studies in gravitational wave astronomy~\citep{Valtonen21,Gao24,Chen18,Dey18,Desai08}. Optical observations for this object go back to the 19th century~\citep{Sillanpaa,Hudec}. This object has been known to emit quasi-periodic outbursts with a period of approximately 12 years. These optical bursts are emitted due to the enhanced accretion from the secondary black hole onto the primary~\citep{Dey19,Valtonen21,Valtonen23}.

In a recent work,~\citet{Gupta23} reported results related to optical flux variability and polarization measurements using a large network of telescopes (Perkins telescope, 1.5~m ``Kanata'' telescope, St Petersburg University 70-cm AZT-8 telescope and 40-cm LX-200 telescope, 0.6~m telescope at the Belogradchik Observatory, Steward Observatory telescope) spanning eight observing seasons from September 2015 to October 2023. These observations are part of a long-term multiwavelength campaign of OJ 287~\citep{Gupta23}.
Observations were presented for eight segments for the above duration, where each segment roughly corresponds to one observing cycle. The third segment was further subdivided into two sub-segments labelled as 3A and 3B because of some peculiar variability features in the third segment. Most recently, ~\citet{Valtonen24} also reported a huge optical flare in November 2021 in both $I$-band and $R$-band with TESS and the prompt5 telescope in Chile as part of the Krakow quasar monitoring campaign, which falls during segment 7. Also, 
TeV gamma-ray emission has also been detected from OJ287 using VERITAS between 1 February and 4 February 2017 (UTC), which falls during Segment 2~\citep{Veritas}, although no details on the duration of the TeV flare are available. Finally this object has also been monitored in X-rays part of the MOMO project~\citep{MOMO}.

In this work, we search for gamma-ray emission during the eight aforementioned segments from 2015-2023 using the Fermi Large Area Telescope (LAT) gamma-ray telescope. The LAT is one of the two instruments onboard this detector. Fermi-LAT is sensitive to high-energy gamma rays from a plethora of astrophysical sources~\citep{Ajello19}. It is a pair-conversion telescope with a field of view of about 2.4 sr that is sensitive to photons between the energy range of 30 MeV to around 300 GeV and has been operating continuously since 2008~\citep{Atwood09}. Fermi-LAT has been monitoring OJ287 
since the start of the mission. OJ 287 has been included in the 3FGL~\citep{3FGL} and 4FGL~\citep{4FGL} catalogs and has been referred to as J0854.8+2006 in the 4FGL~\citep{Ajello}. 
Previous gamma-ray variability studies with Fermi-LAT have been reported in a number of works~\citep{Abdo09,Neronov,Kushwaha13,Prince,Komossa22}. ~\citet{Prince} found that the isotropic gamma-ray luminosity to be between $10^{47}-10^{48}$ ergs/sec. The average and peak photon flux were found to be $\sim 2.65$ and $\sim 10$ $\times 10^{-8} \rm{ph. cm^{-2} sec^{-1}}$ in the energy range between 0.1-300 GeV~\citep{Prince}. This analysis found a variation of a factor of five between the lowest and highest flux states. We now present the results from our analysis of Fermi-LAT data between 2015-2023.

This manuscript is structured as follows. Our analysis of the Fermi-LAT data is discussed in Sect~\ref{sec:analysis}. Comparison with other multi-wavelength flares for OJ 287 is discussed in Sect.~\ref{sec:opticalgammacomp}. We conclude in Sect.~\ref{sec:conclusions}.

\section{Analysis of Fermi-LAT data}
\label{sec:analysis}
We analyzed the Fermi-LAT data using the {\tt EasyFermi} software~\citep{easyFermi}, which is based on both {\tt FermiTools} and {\tt Fermipy} packages~\citep{Fermipy}. 
The Fermi-LAT data from 2015 - 2023 was divided into eight time segments. For each segment, we searched for OJ 287 in the {\tt 4FGL-DR3} catalog, within an energy range of 100 MeV to 300 GeV (unless specified) and a search radius of $7.5^{\circ}$. We used 20-time bins for light curves for Segments 1, 2, 3A, 5, 7, and 8; 8 bins for Segment 3B; 7 bins for Segment 4; and 10 bins for Segment 6 (as larger number of bins have resulted in null data points in some bins). We then proceeded to fit the Spectral Energy Distribution (SED) using a simple power law model and also a log-parabola model. The power-law spectrum is defined as follows:
\begin{equation}
dN/dE = N_0 \left(E/E_0\right)^{-\alpha},
\label{eq:1}
\end{equation}
where $\alpha$ is the spectral index, $N_0$ is the normalization and $E_0$ is the pivot energy. Similarly, the log-parabola spectrum is defined according to:
\begin{equation}
 dN/dE = N_0 \left( \frac{E}{E_0} \right) ^ { - \alpha - \beta \log{ \left( \frac{E}{E_0} \right) } }
 \label{eq:logparabola}
\end{equation}


\section{Results}
We now show the results for the gamma-ray light curves and SED, with and without correction for absorption due to extragalactic background light, and photon flux light curves for OJ 287 during the aforementioned eight segments. A tabular summary of the results is summarized in Table~\ref{tab:s2:pars}, where we show the test statistic (TS)~\citep{Mattox,Pasumarti, Manna}, photon flux, energy flux, and the isotropic gamma-ray luminosity, which has been calculated using Equation 5 of ~\citet{Prince21}. The best-fit spectral parameters for both the power-law and log-parabola model can be found in Table~\ref{tab:my_label}. We also confirmed that that the value of ``Fit Quality'' reported by {\tt EasyFermi} is equal to three for all the analyzed segments, indicating the fit is excellent and there is no problem in the convergence of the log likelihood~\cite{easyFermi}.

\subsection{Segment 1 (JD: 2457284 - 2457536)}
\begin{figure}[H]
 \centering
 \includegraphics[width=0.5\textwidth]{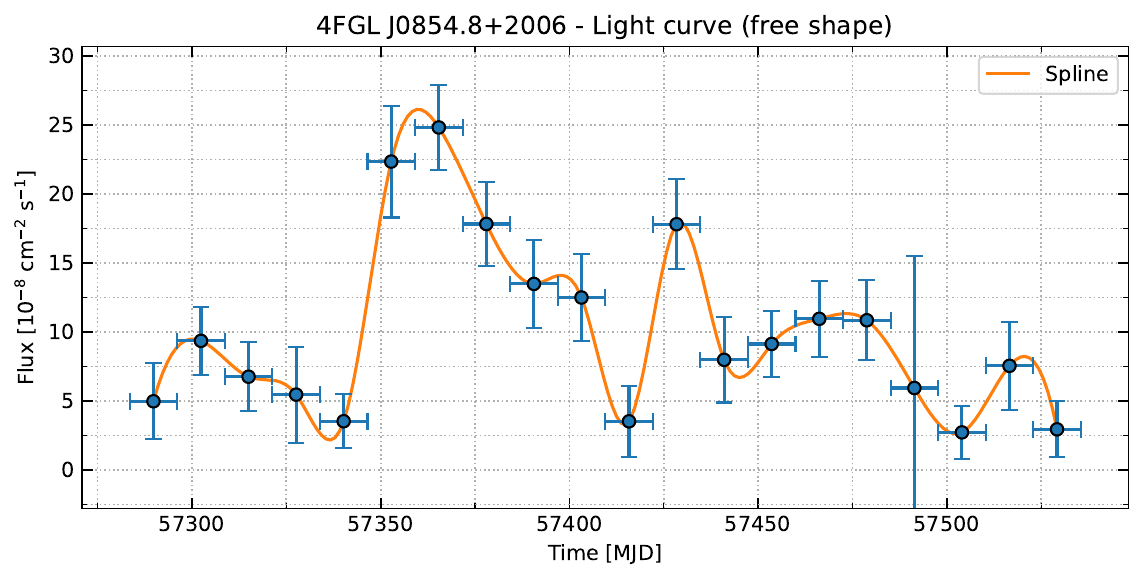}
 \caption{Segment 1 (MJD : 	57283.5	-	57535.5	) - Light Curve; binwidth = 12 days.}
 \label{fig:s1:LC}
\end{figure}

\begin{figure}[H]
 \centering
 \includegraphics[width=0.5\textwidth]{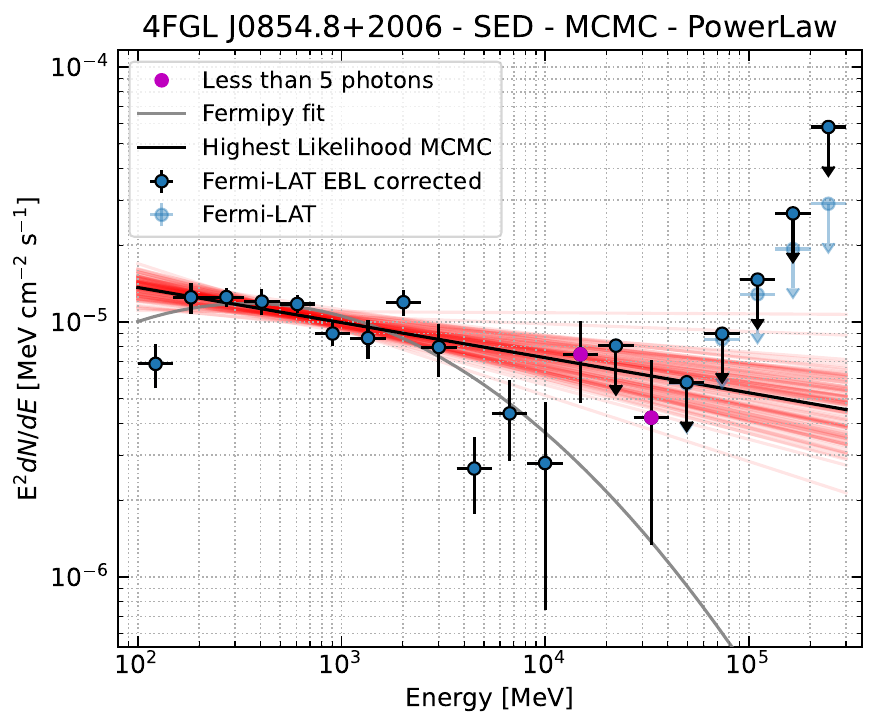}
 \caption{Segment 1 - SED. The black solid line shows the power-law fit while the gray line shows the log-parabola fit from Fermipy. }
 \label{fig:s1:SED}
\end{figure}

\begin{figure}[H]
\centering
\includegraphics[width=0.4\textwidth]{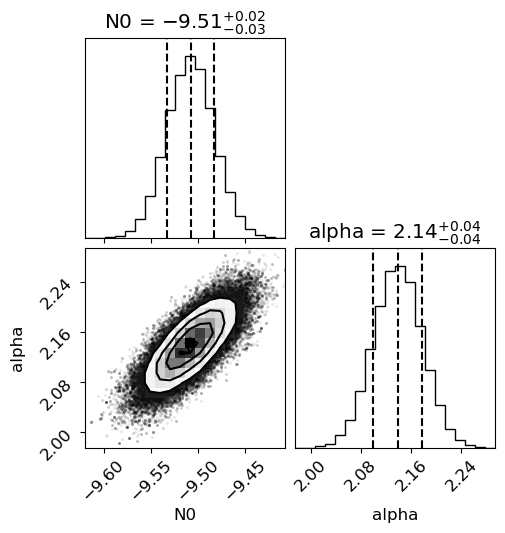}
 \caption{Segment 1 - SED Best fit parameters: $\log_{10} N_0 = -9.51 \pm 0.03$ ; $\alpha = 2.10 \pm 0.04$; $E_0 = 636.6$ MeV. 
 }
 \label{fig:s1:SED:par}
\end{figure}

The photon flux light curve (sampled at a cadence of 12 days) for segment 1 and the corresponding SED can be found in Fig.~\ref{fig:s1:LC} and ~\ref{fig:s1:SED}, respectively. The average photon and energy flux are given by $ (1.08 \pm 0.07) \times 10^{-7} \rm{cm^{-2} s^{-1}}$ and $(4.65 \pm 0.27) \times 10^{-5}$ $\rm{MeV~cm^{-2} s^{-1}}$ respectively. The light curve shows a variability upto a factor of five between the smallest and largest flux. The SED shows non-zero values up to 20 GeV and upper limits beyond that. The marginalized posteriors for the spectral fit parameters can be found in Fig.~\ref{fig:s1:SED:par}. 

\vspace{2cm}
\subsection{Segment 2 (JD: 2457645 - 2457920)}
\begin{figure}[h]
 \centering
 \includegraphics[width=0.5\textwidth]{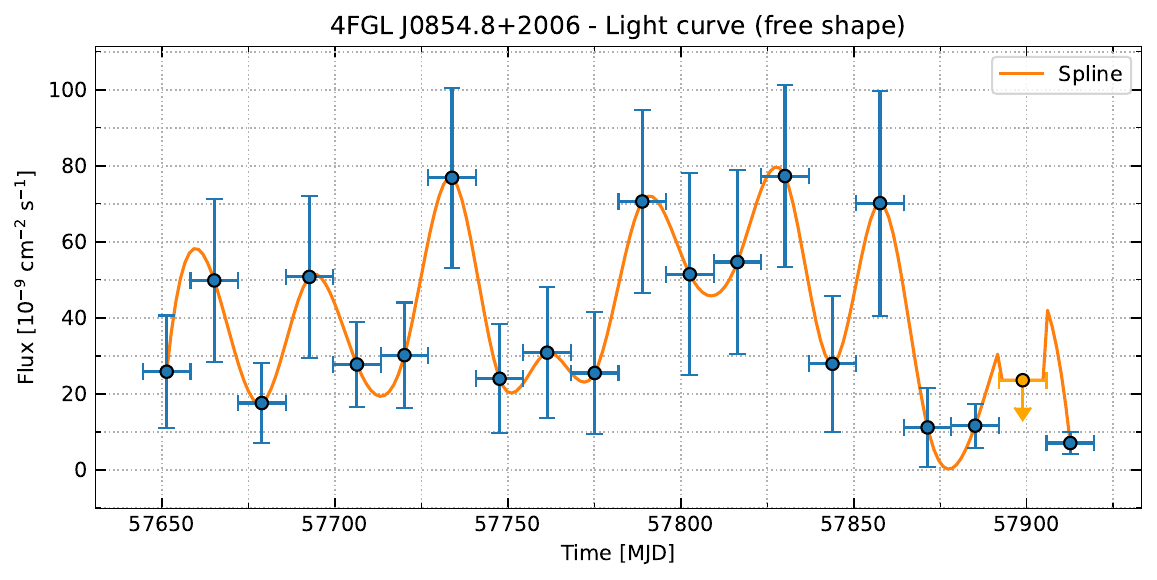}
 \caption{Segment 2 (MJD : 	57644.5	-	57919.5	) - Light Curve; \rthis{binwidth = 13.75 days}}
 \label{fig:s2:LC}
\end{figure}

\begin{figure}[h]
 \centering
 \includegraphics[width=0.5\textwidth]{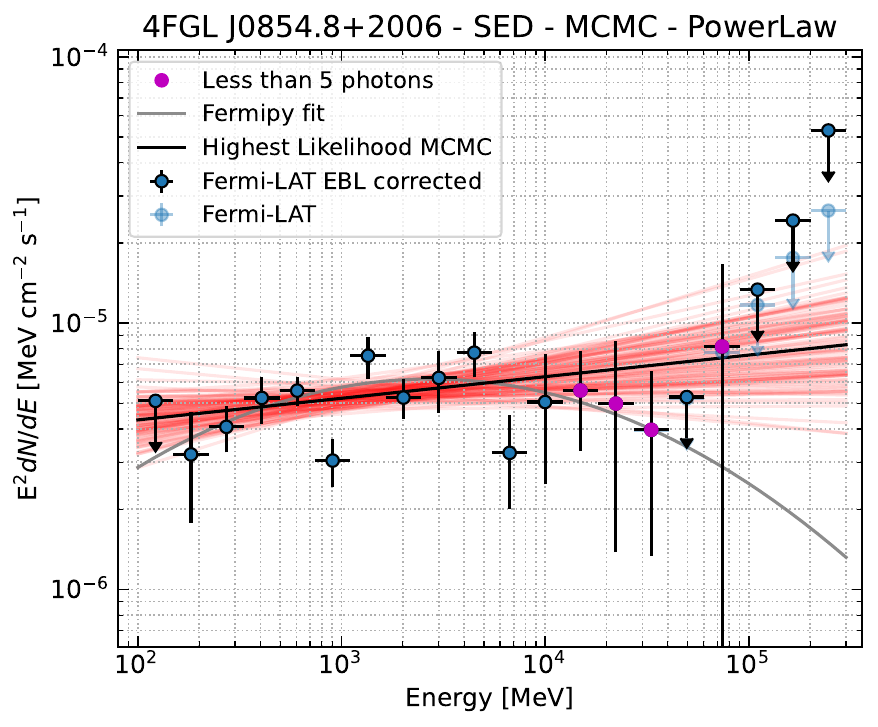}
 \caption{Segment 2 - SED. The black solid line shows the power-law fit while the gray line shows the log-parabola fit from Fermipy.}
 \label{fig:s2:SED}
\end{figure}

\begin{figure}[H]
 \centering
 \includegraphics[width=0.4\textwidth]{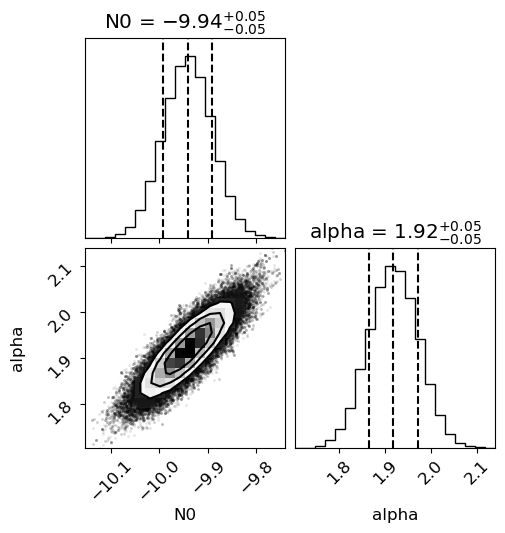}
 \caption{Segment 2 - SED Best fit parameters: $\log_{10} N_0 = -9.94 \pm 0.05$ ; $\alpha = 1.92 \pm 0.05$ ; $E_0 = 720.59$ MeV.
 }
 \label{fig:s2:SED:par}
\end{figure}

The photon flux light curve, as well as the SED for segment 2, can be found in Fig.~\ref{fig:s2:LC} and ~\ref{fig:s2:SED}, respectively. The average photon and energy flux are given by $(4.07 \pm 0.05) \times 10^{-8} \rm{cm^{-2} s^{-1}}$ and $(3.52 \pm 0.09) \times 10^{-5} \rm{MeV~cm^{-2} s^{-1}}$ respectively. 
Here, we see a factor of four variation between the smallest and largest flux values. The posteriors for the best-fit power law parameters can be found in Fig.~\ref{fig:s2:SED:par}. The SED shows non-zero flux up to 10 GeV and upper limits beyond that.

\vspace{2cm}
\subsection{Segment 3A (JD: 2458016 - 2458142)}
\begin{figure}[h]
 \centering
 \includegraphics[width=0.5\textwidth]{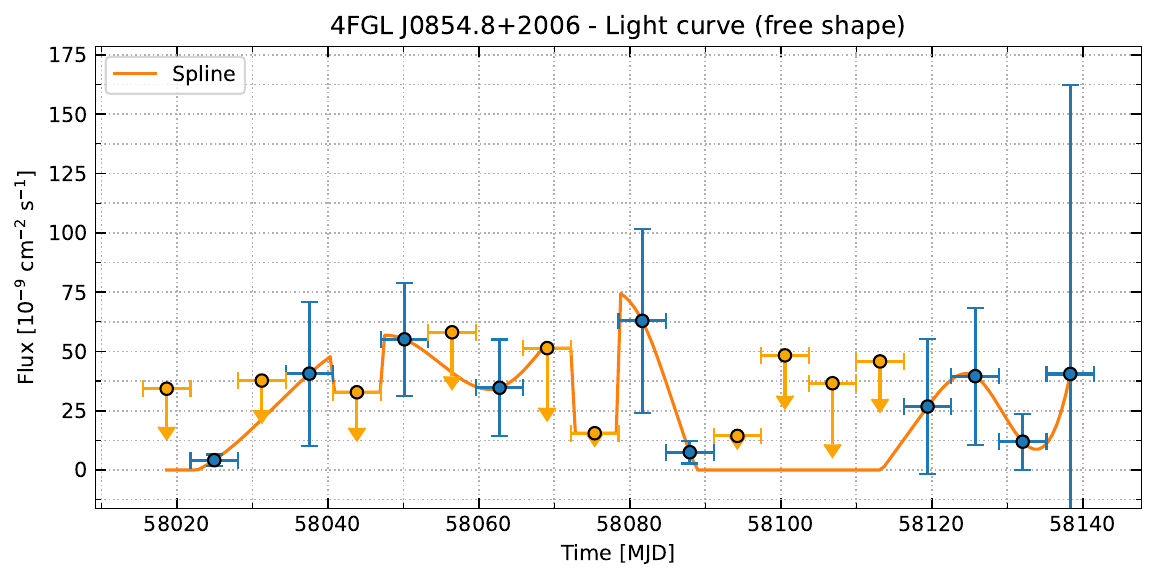}
 \caption{Segment 3A (MJD : 	58015.5	-	58141.5	) - Light Curve; \rthis{binwidth = 6.3 days}}
 \label{fig:s3A:LC}
\end{figure}

\begin{figure}[h]
 \centering
 \includegraphics[width=0.5\textwidth]{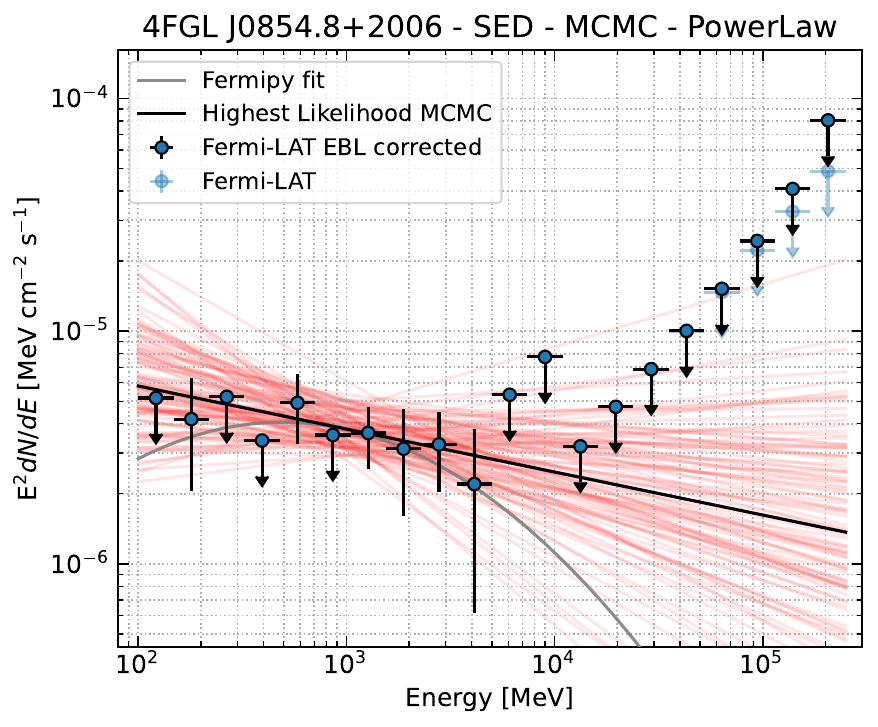}
 \caption{Segment 3A - SED. The black solid line shows the power-law fit while the gray line shows the log-parabola fit from Fermipy.}
 \label{fig:s3A:SED}
\end{figure}

\begin{figure}[H]
 \centering
 \includegraphics[width=0.4\textwidth]{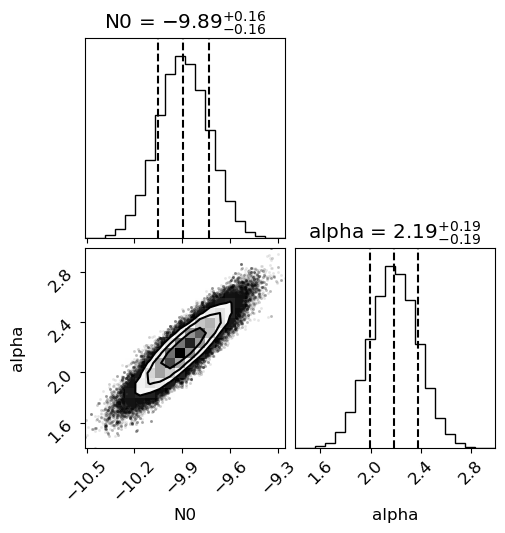}
 \caption{Segment 3A - SED Best fit parameters: $\log_{10} N_0 = -9.89 \pm 0.16$ ; $\alpha = 2.19 \pm 0.19$ ; $E_0 = 846.77$ MeV.
 }
 \label{fig:s3A:SED:par}
\end{figure}

The photon flux light curve and SED for segment 3A can be found in Fig.~\ref{fig:s3A:LC} and ~\ref{fig:s3A:SED}, respectively. The average photon and energy flux are given by $(3.44 \pm 0.95) \times \rm{10^{-8} cm^{-2} s^{-1}}$ and $ (1.53 \pm 0.26) \times \rm{10^{-5} MeV~cm^{-2} s^{-1}}$ respectively. 
Here, we see a nearly constant flux within $1\sigma$ error bars. The SED shows non-zero flux up to 4 GeV and upper limits beyond that. The posteriors for the best-fit power law parameters can be found in Fig.~\ref{fig:s3A:SED:par}.

\vspace{2cm}
\subsection{Segment 3B (JD: 2458156 - 2458292)}
\begin{figure}[h]
 \centering
 \includegraphics[width=0.5\textwidth]{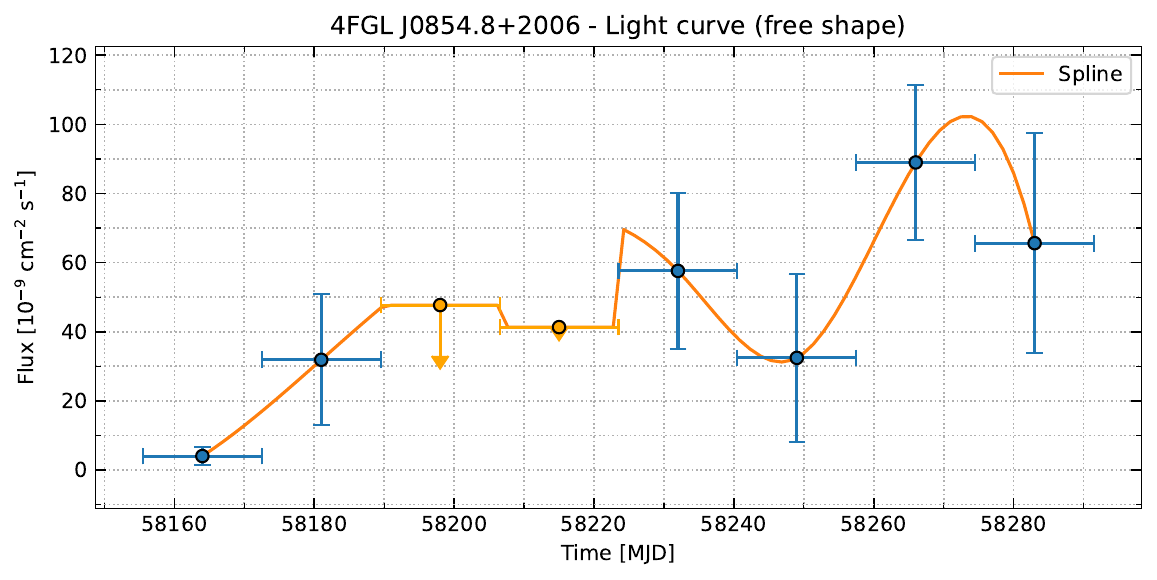}
 \caption{Segment 3B (MJD : 	58155.5	-	58291.5	) - Light Curve ; \rthis{binwidth = 17 days}}
 \label{fig:s3B:LC}
\end{figure}

\begin{figure}[h]
 \centering
 \includegraphics[width=0.5\textwidth]{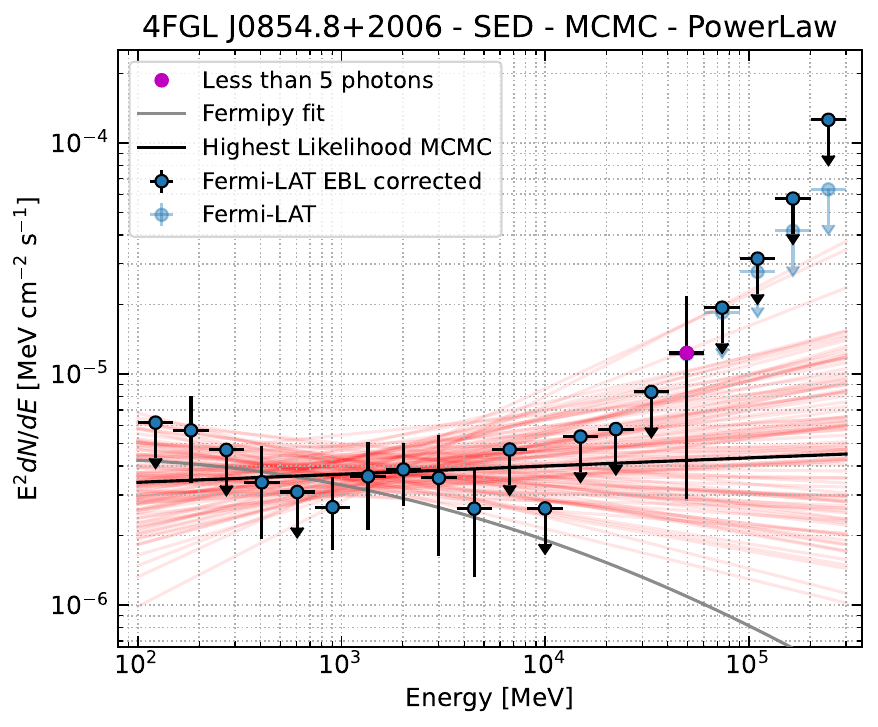}
 \caption{Segment 3B - SED. \rthis{The black solid line shows the power-law fit while the gray line shows the log-parabola fit from Fermipy.}}
 \label{fig:s3B:SED}
\end{figure}

\begin{figure}[H]
 \centering
 \includegraphics[width=0.4\textwidth]{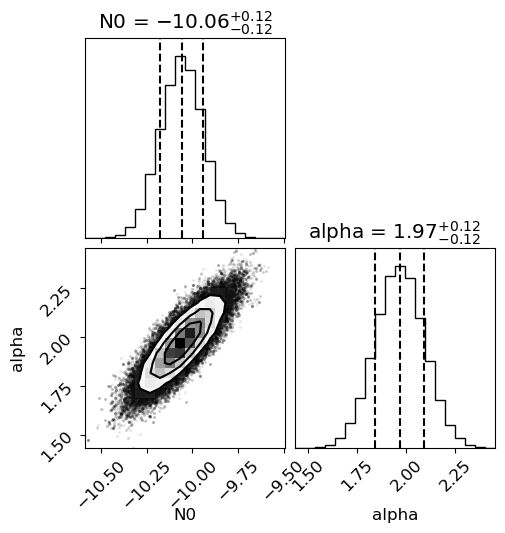}
 \caption{Segment 3B - SED Best fit parameters: $\log_{10} N_0 = -10.06 \pm 0.12$ ; $\alpha = 1.97 \pm 0.12$ ; $E_0 = 803.72$ MeV. 
 }
 \label{fig:s3B:SED:par}
\end{figure}

The photon flux light curve and SED for segment 3B can be found in Fig.~\ref{fig:s3B:LC} and ~\ref{fig:s3B:SED}, respectively. The average photon and energy flux are given by $ (3.88 \pm 0.67) \times 10^{-8} \rm{cm^{-2} s^{-1}}$ and $ (1.86 \pm 0.33) \times 10^{-5} \rm{MeV~cm^{-2} s^{-1}}$ respectively. From the SED, we find non-zero flux until 4 GeV and upper limits beyond that, except for one data point near 50 GeV. Similar to segment 3A, we see an approximately constant flux, except for the first data point. The posteriors for the best-fit spectral parameters can be found in Fig.~\ref{fig:s3B:SED:par}.
\vspace{2cm}
\subsection{Segment 4 (JD: 2458380 - 2458632)}
\begin{figure}[h]
 \centering
 \includegraphics[width=0.5\textwidth]{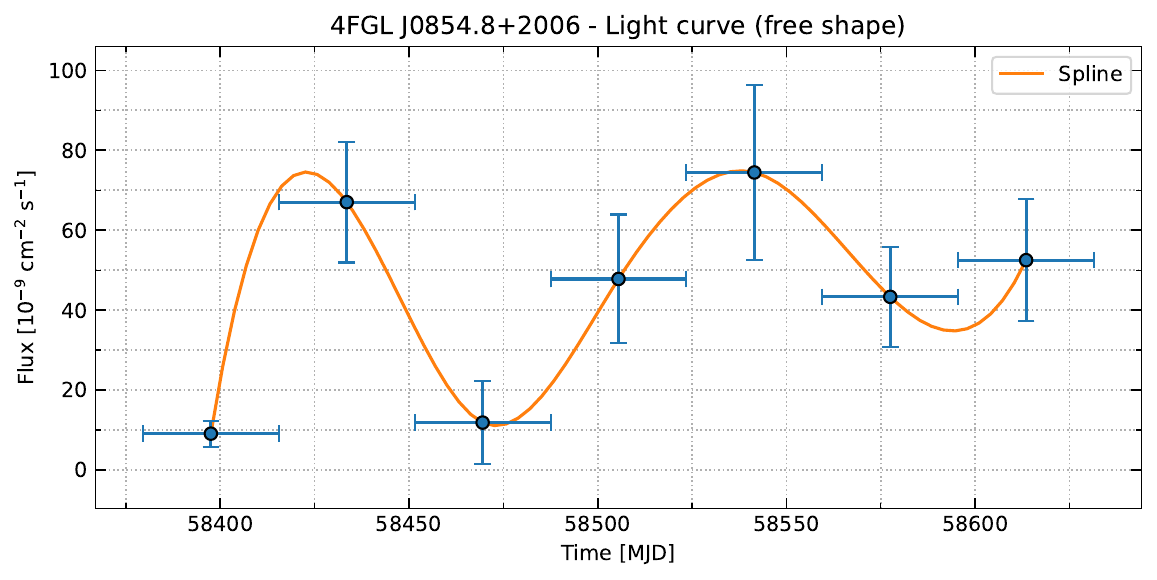}
 \caption{Segment 4 (MJD : 	58379.5	-	58631.5	) - Light Curve ; \rthis{binwidth = 36 days}}
 \label{fig:s4:LC}
\end{figure}

\begin{figure}[h]
 \centering
 \includegraphics[width=0.5\textwidth]{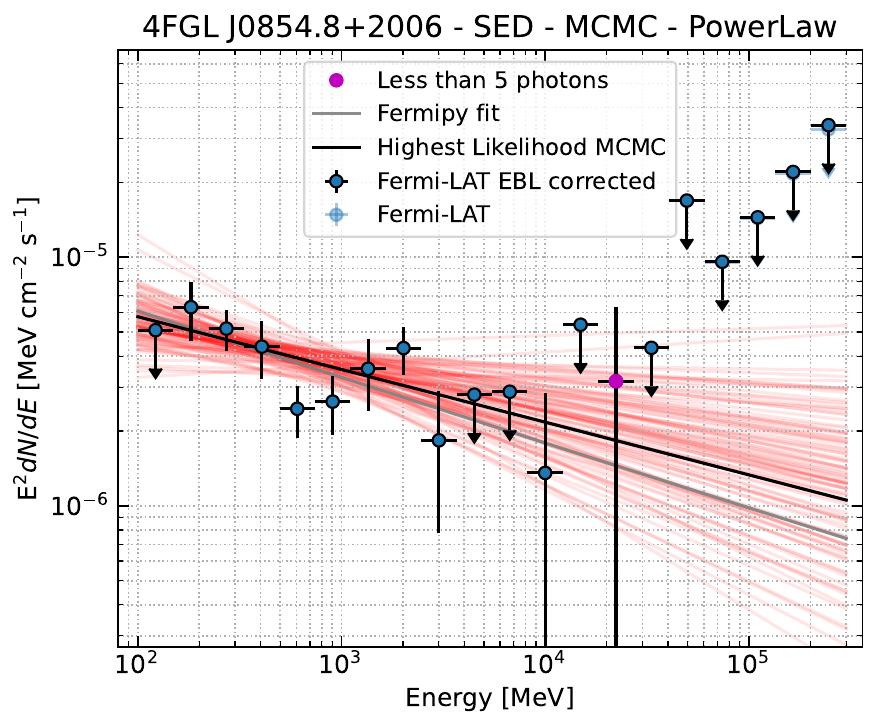}
 \caption{Segment 4 - SED. The black solid line shows the power-law fit while the gray line shows the log-parabola fit from Fermipy.}
 \label{fig:s4:SED} 
 \end{figure}

\begin{figure}[H]
 \centering
 \includegraphics[width=0.4\textwidth]{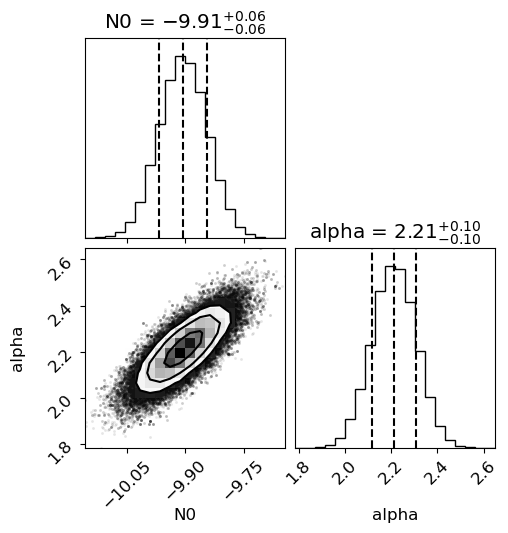}
 \caption{Segment 4 - SED Best fit parameters: $\log_{10} N_0 = -9.90 \pm 0.06$ ; $\alpha = 2.21 \pm 0.10$ ; $E_0 = 545.92$ MeV. 
 }
 \label{fig:s4:SED:par}
\end{figure}

The photon flux light curve and SED for segment 4 can be found in Fig.~\ref{fig:s4:LC} and ~\ref{fig:s4:SED}, respectively. The average photon and energy flux are given by $ (4.80 \pm 0.78) \times 10^{-8} \rm{cm^{-2} s^{-1}}$ and $ (2.02 \pm 0.34) \times 10^{-5} \rm{MeV~cm^{-2} s^{-1}}$ respectively. 
From the SED, we find non-zero flux until 3 GeV and upper limits beyond that. Similar to segment 3A, we see an approximately constant flux, except for the first data point. The light curve also shows a factor of five variation between the minimum and maximum flux. The posteriors for the best-fit spectral parameters can be found in Fig.~\ref{fig:s4:SED:par}.
\vspace{2cm}
\subsection{Segment 5 (JD: 2458755 - 2459009)}
\begin{figure}[h]
 \centering
 \includegraphics[width=0.5\textwidth]{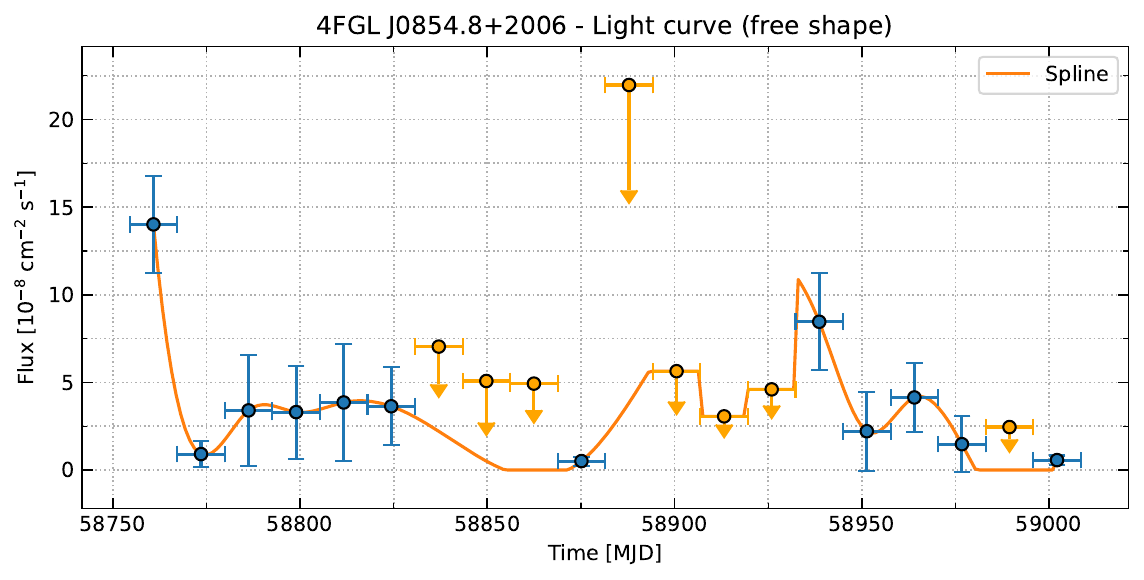}
 \caption{Segment 5 (MJD : 	58754.5	-	59008.5	) - Light Curve ; binwidth = 12.7 days}
 \label{fig:s5:LC}
\end{figure}

\begin{figure}[h]
 \centering
 \includegraphics[width=0.5\textwidth]{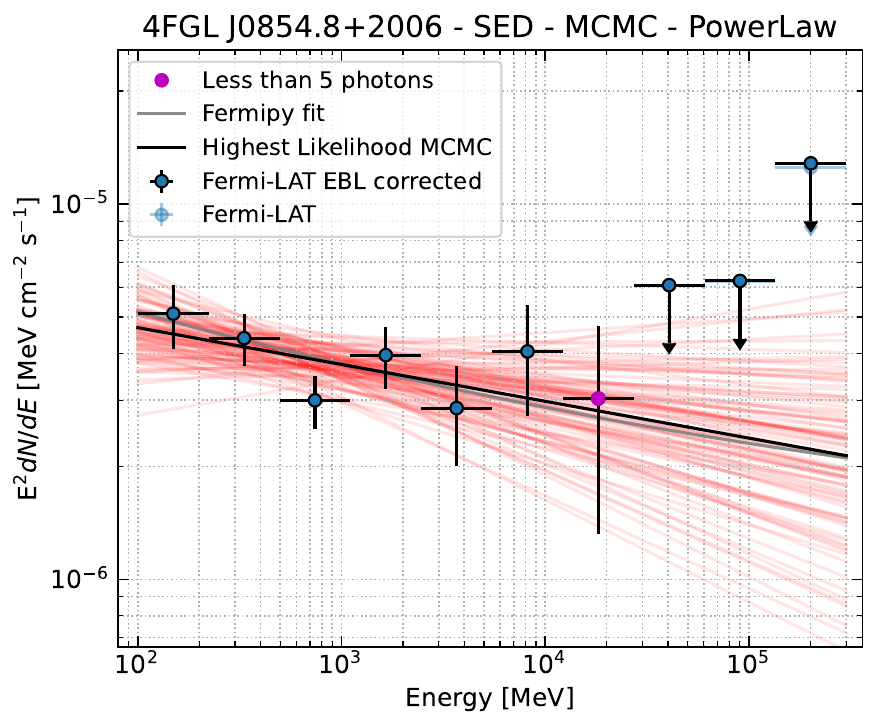}
 \caption{Segment 5 - SED. \rthis{The black solid line shows the power-law fit while the gray line shows the log-parabola fit from Fermipy.}}
 \label{fig:s5:SED}
\end{figure}

\begin{figure}[H]
 \centering
 \includegraphics[width=0.39\textwidth]{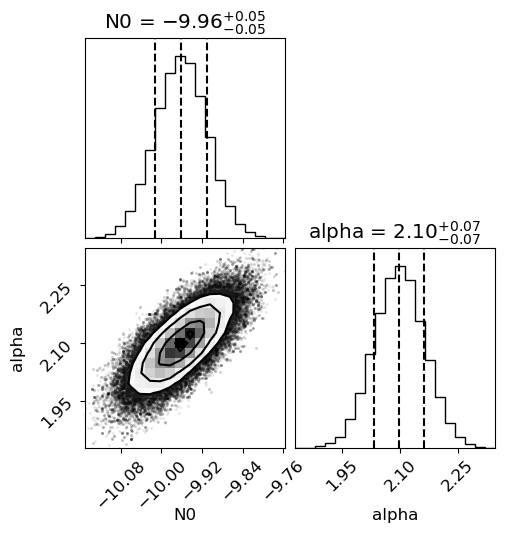}
 \caption{Segment 5 - SED Best fit parameters: $\log_{10} N_0 = -9.96 \pm 0.05$ ; $\alpha = 2.10^{+0.06}_{-0.07}$ ; $E_0 = 921.45$ MeV.
 }
 \label{fig:s5:SED:par}
\end{figure}

The photon flux light curve and SED for segment 5 can be found in Fig.~\ref{fig:s5:LC} and ~\ref{fig:s5:SED}, respectively. The average photon and energy flux are given by $ (4.52 \pm 0.72) \times 10^{-8} \rm{cm^{-2} s^{-1}}$ and $ (2.60 \pm 0.45) \times 10^{-5} \rm{MeV~cm^{-2} s^{-1}}$ respectively. 
From the SED, we find non-zero flux until 20 GeV and upper limits beyond that. Similar to segment 3A, we see an approximately constant flux, except for the first data point. The light curve steeply falls near the beginning of the segment (within a factor of 15) and then asymptotes to a nearly constant value for the remainder of the segment. The posteriors for the best-fit spectral parameters can be found in Fig.~\ref{fig:s5:SED:par}.
\vspace{2cm}
\subsection{Segment 6 (JD: 2459108 - 2459338)}
\begin{figure}[h]
 \centering
 \includegraphics[width=0.5\textwidth]{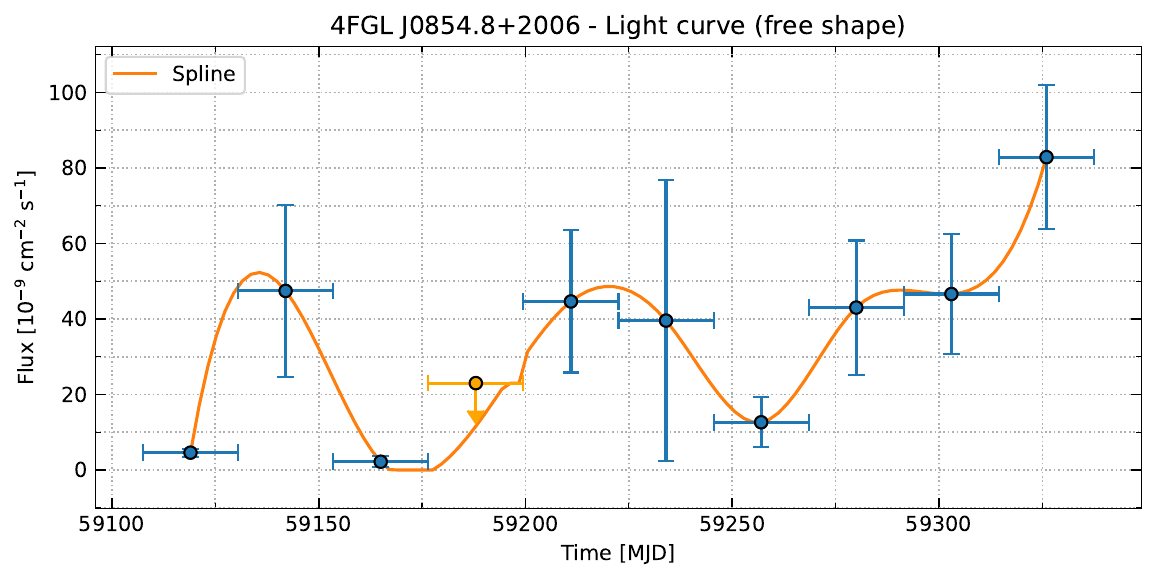}
 \caption{Segment 6 (MJD : 	59107.5	-	59337.5	) - Light Curve ; binwidth = 23 days}
 \label{fig:s6:LC}
\end{figure}

\begin{figure}[h]
 \centering
 \includegraphics[width=0.5\textwidth]{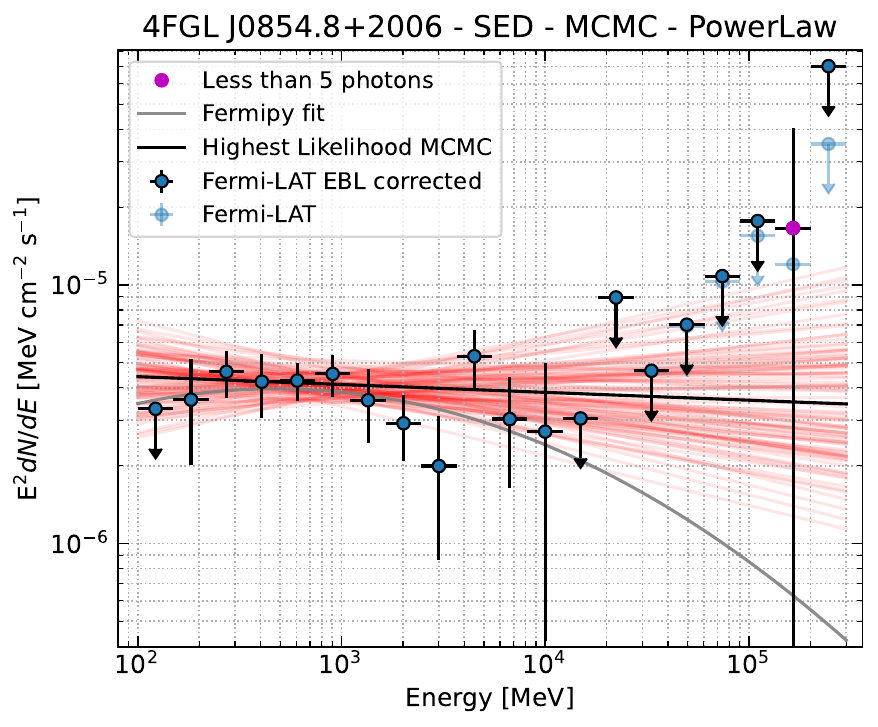}
 \caption{Segment 6 - SED. \rthis{The black solid line shows the power-law fit while the gray line shows the log-parabola fit from Fermipy.}}
 \label{fig:s6:SED}
\end{figure}

The photon flux light curve and SED for segment 6 can be found in Fig.~\ref{fig:s6:LC} and ~\ref{fig:s6:SED}, respectively. The average photon and energy flux are given by $(3.72 \pm 0.67) \times 10^{-8} \rm{cm^{-2} s^{-1}}$ and $(2.06 \pm 0.33) \times 10^{-5} \rm{MeV~cm^{-2} s^{-1}}$ respectively. 
 
\begin{figure}[H]
 \centering
 \includegraphics[width=0.4\textwidth]{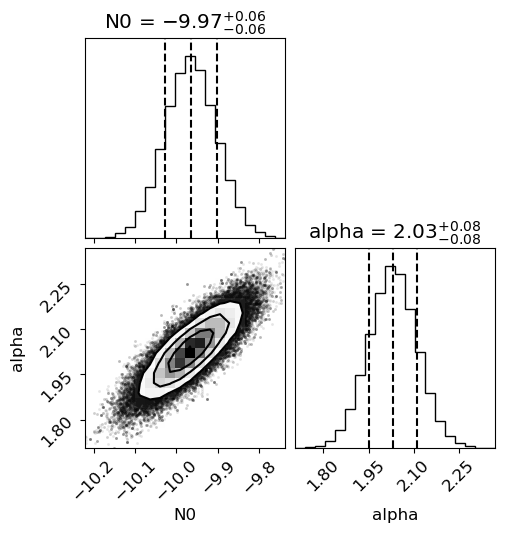}
 \caption{Segment 6 - SED Best fit parameters: $\log_{10} N_0 = -9.96 \pm 0.06$ ; $\alpha = 2.04 \pm 0.08$ ; $E_0 = 745.69$ MeV.
 }
 \label{fig:s6:SED:par}
\end{figure}

From the SED, we find non-zero flux up to approximately 10 GeV and upper limits beyond that. Similar to segment 3A, we see an approximately constant flux, except for the first data point. The light curve mainly shows a constant flux, except for the first and third points, which depict very small values and then rise towards the end of the segment. The posteriors for the best-fit spectral parameters can be found in Fig.~\ref{fig:s6:SED:par}.
\vspace{2cm}
\subsection{Segment 7 (JD: 2459520 - 2459750)}
\begin{figure}[h]
 \centering
 \includegraphics[width=0.5\textwidth]{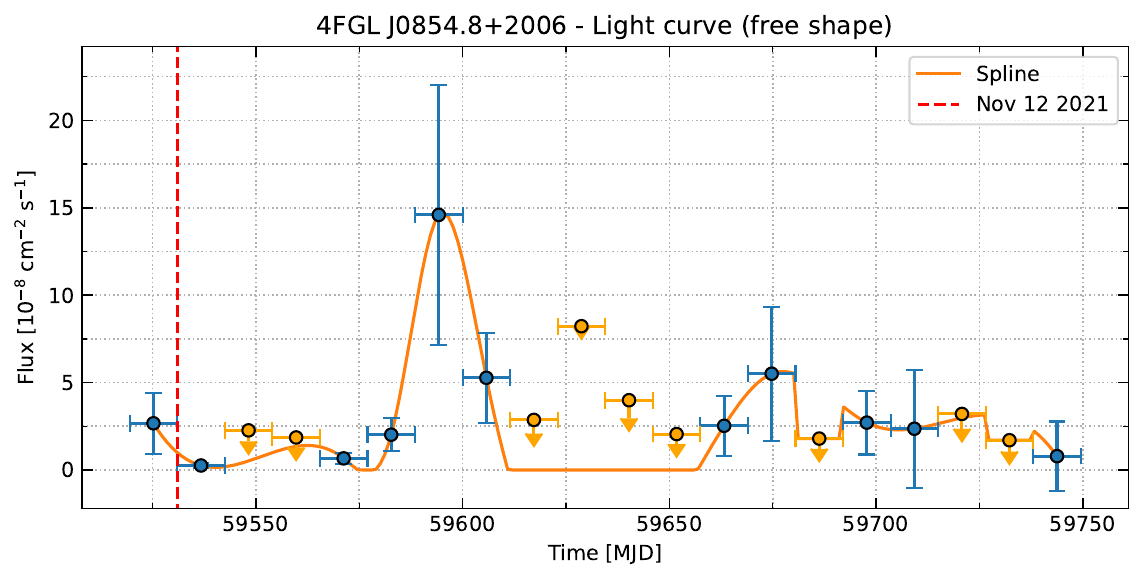}
 \caption{Segment 7 (MJD : 	59519.5	-	59749.5	) - Light Curve ; binwidth = 11.5 days. The dashed vertical line corresponds to the start of the optical flare detected in $I$ and $R$ band as reported in ~\citet{Valtonen24}.}
 \label{fig:s7:LC}
\end{figure}

\begin{figure}[h]
 \centering
 \includegraphics[width=0.5\textwidth]{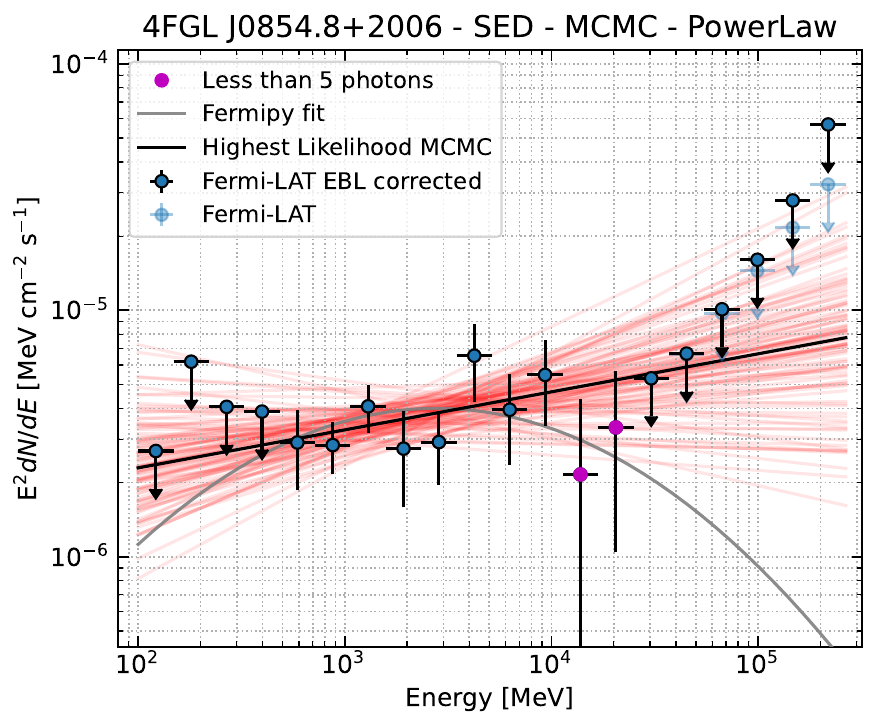}
 \caption{Segment 7 - SED. \rthis{The black solid line shows the power-law fit while the gray line shows the log-parabola fit from Fermipy.}}
 \label{fig:s7:SED}
\end{figure}

\begin{figure}[H]
 \centering
 \includegraphics[width=0.38\textwidth]{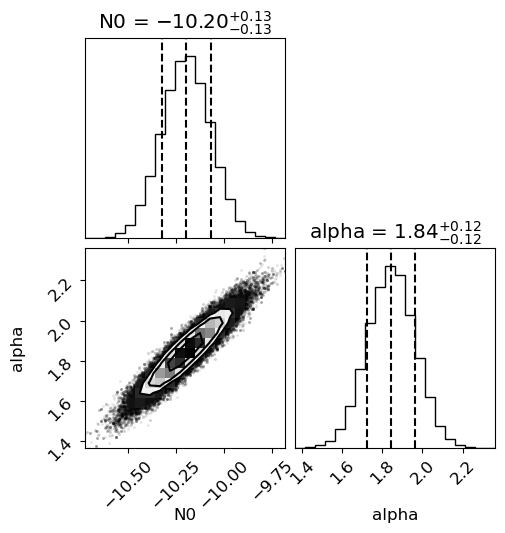}
 \caption{Segment 7 - SED Best fit parameters: $\log_{10} N_0 = -10.20 ^{+0.13}_{-0.12}$ ; $\alpha = 1.85 ^{+0.12}_{-0.11}$ ; $E_0 = 1081.02$ MeV. 
 }
 \label{fig:s7:SED:par}
\end{figure}

The photon flux light curve and SED for segment 7 can be found in Fig.~\ref{fig:s7:LC} and ~\ref{fig:s7:SED}, respectively. The average photon and energy flux are given by $(2.06 \pm 0.25)\times 10^{-8} \rm{cm^{-2} s^{-1}}$ and $(1.95 \pm 0.27) \times 10^{-5}$ $\rm{MeV~cm^{-2} s^{-1}}$ respectively. 
 From the SED, we find non-zero flux up to approximately 20 GeV and upper limits beyond that. Similar to segment 3A, we see an approximately constant flux, except for the first data point. The light curve shows a sinusoidal trend near the start of the segment until MJD of 59600 followed by \rthis{upper limits} for the next 50 days. The light curve then shows a constant value 
 until the end of the segment. We also do not find any correlation between the optical flare in $I$ and $R$ band reported in ~\citet{Valtonen24} and the gamma-ray light curve. The posteriors for the best-fit spectral parameters can be found in Fig.~\ref{fig:s7:SED:par}.
\vspace{2cm}
\subsection{Segment 8 (JD: 2459810-2459999) }
\begin{figure}[h]
 \centering
 \includegraphics[width=0.5\textwidth]{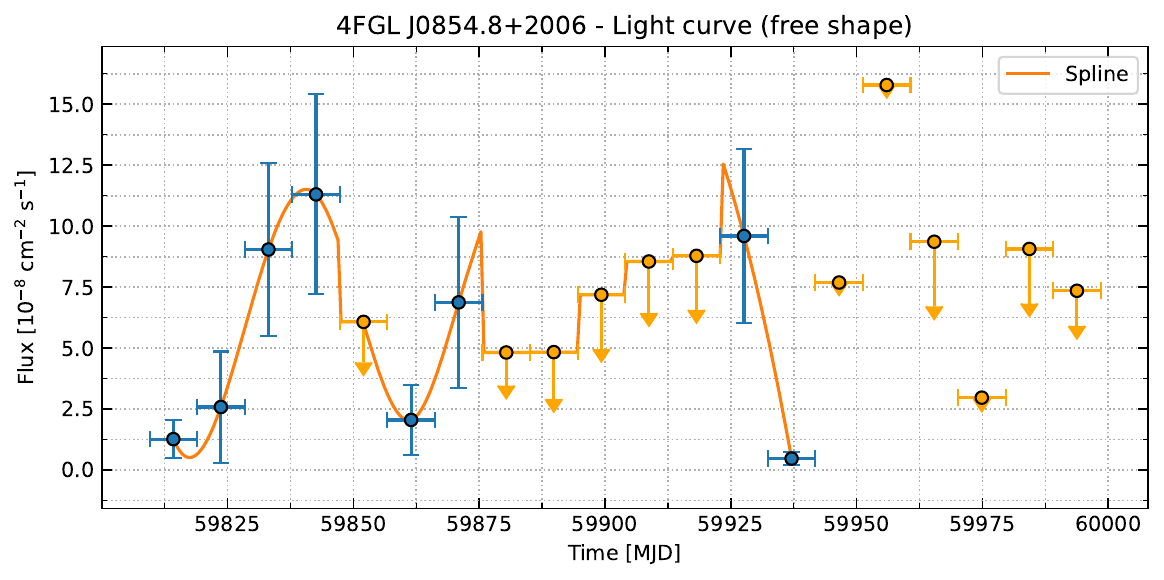}
 \caption{Segment 8 (MJD : 	59809.5	-	59998.5	) - Light Curve ; binwidth = 9.45 days.}
 \label{fig:s8:LC}
\end{figure}

\begin{figure}[h]
 \centering
 \includegraphics[width=0.5\textwidth]{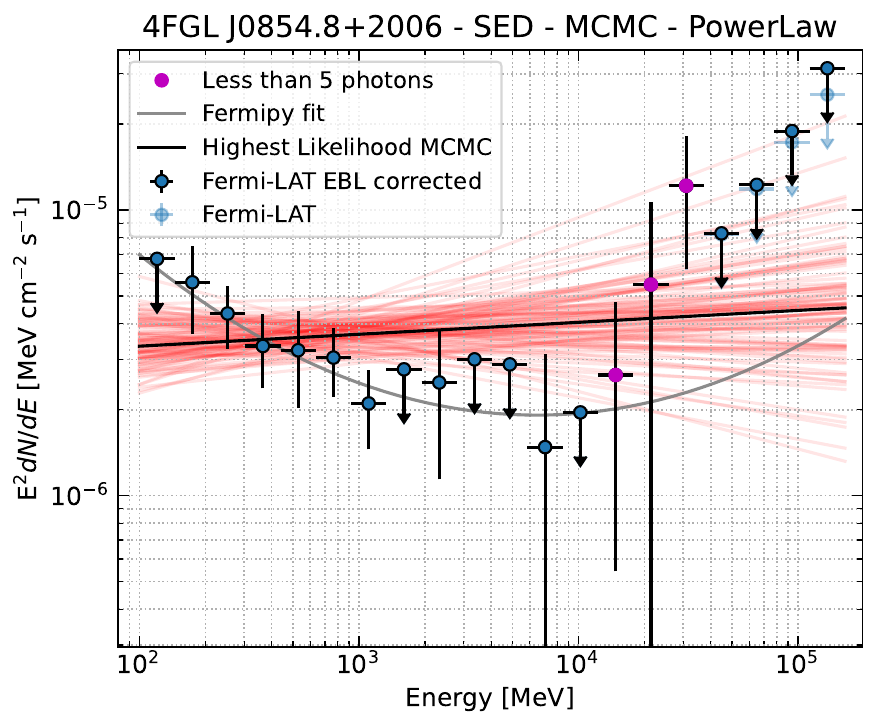}
 \caption{Segment 8 - SED. \rthis{The black solid line shows the power-law fit while the gray line shows the log-parabola fit from Fermipy.}}
 \label{fig:s8:SED}
\end{figure}

\begin{figure}[H]
 \centering
 \includegraphics[width=0.4\textwidth]{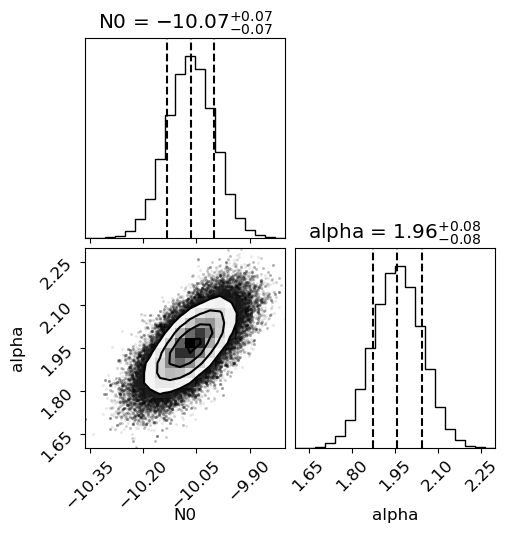}
 \caption{Segment 8 - SED Best fit parameters: $\log_{10} N_0 = -10.07 \pm 0.07$ ; $\alpha = 1.96 \pm 0.08$ ; $E_0 = 705.82$ MeV.
 }
 \label{fig:s8:SED:par}
\end{figure}

The photon flux light curve and SED for segment 8 can be found in Fig.~\ref{fig:s8:LC} and ~\ref{fig:s8:SED}, respectively. We note that the optical data was found to be in a low flux state in ~\citet{Gupta23}. The average photon and energy flux are given by $(4.61 \pm 0.18) \times 10^{-8} \rm{cm^{-2} s^{-1}}$ and $(2.15 \pm 0.14) \times 10^{-5} \rm{MeV~cm^{-2} s^{-1}}$ respectively. 
We find that only six data points in the light curve have non-zero values, with the rest showing upper limits. Similarly, the SED only shows values until 2~GeV.
The posteriors for the best-fit spectral parameters can be found in Fig.~\ref{fig:s8:SED:par}.

\vspace{2cm}
\subsection{Segments 1 - 8 (JD: 2457284 - 2459999)}
\begin{figure}[h]
 \centering
 \includegraphics[width=0.5\textwidth]{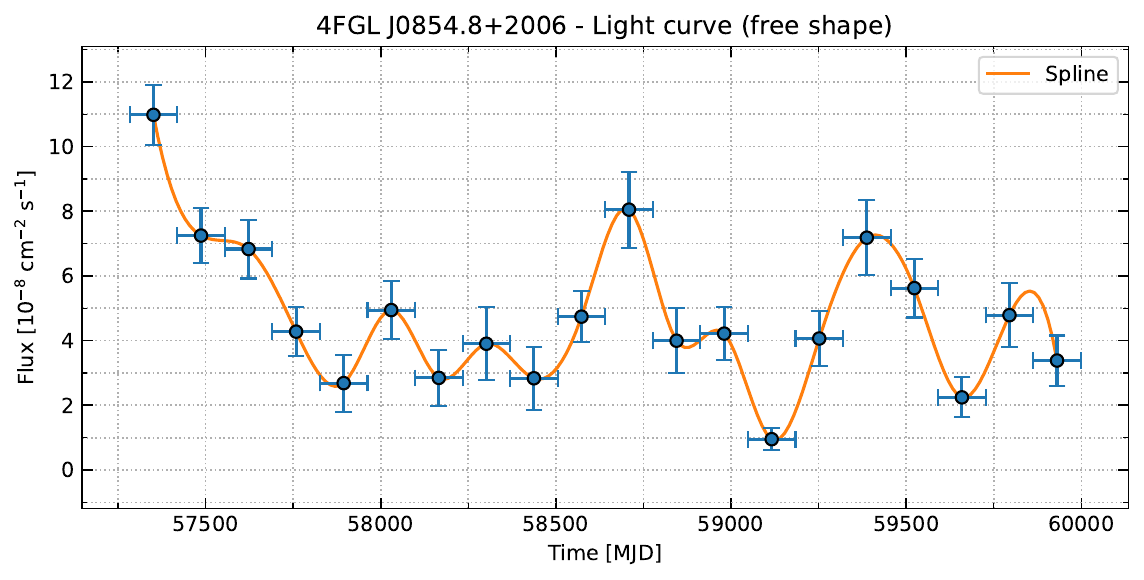}
 \caption{Segments 1-8 (MJD : 	57283.5	-	59998.5) - Light Curve ; binwidth = 135.75 days.}
 \label{fig:all:LC}
\end{figure}

\begin{figure}[h]
 \centering
 \includegraphics[width=0.5\textwidth]{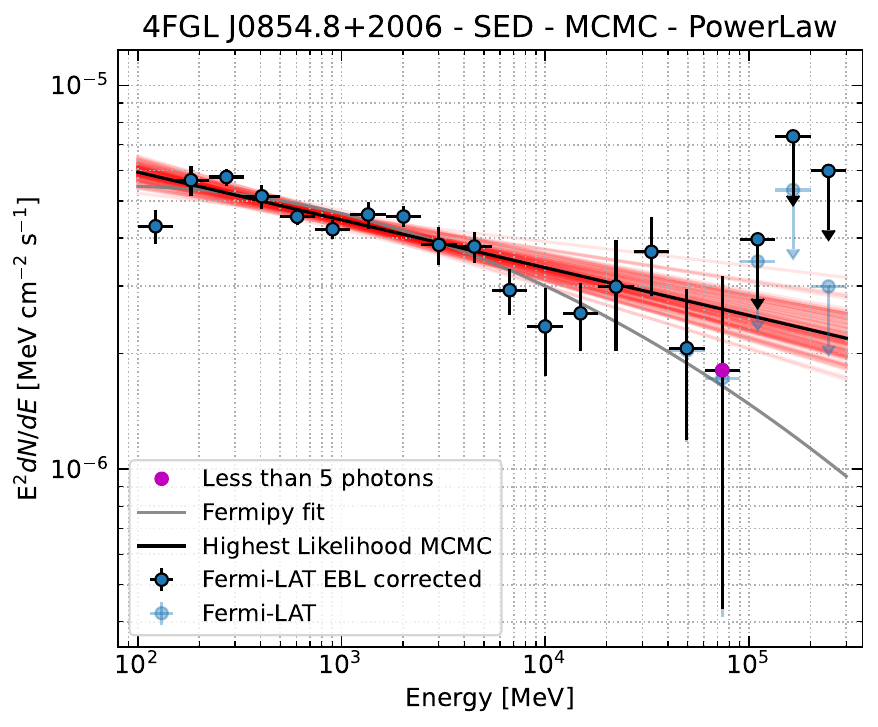}
 \caption{Segments 1-8 - SED. The black solid line shows the power-law fit while the gray line shows the log-parabola fit from Fermipy.}
 \label{fig:all:SED}
\end{figure}

\begin{figure}[H]
 \centering
 \includegraphics[width=0.4\textwidth]{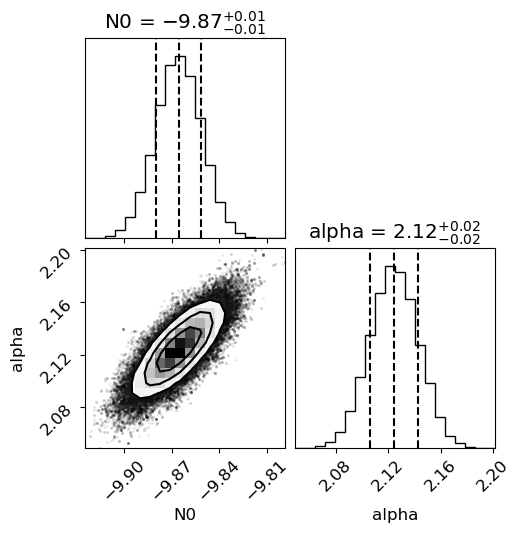}
 \caption{Segments 1-8 - SED Best fit parameters: $\log_{10} N_0 = -9.87 \pm 0.01$ ; $\alpha = 2.12 \pm 0.02$ ; $E_0 = 715.56$ MeV.
 }
 \label{fig:all:SED:par}
\end{figure}

We now do a holistic analysis of the entire data for all eight segments. 
The photon flux light curve and SED for segment 8 can be found in Fig.~\ref{fig:all:LC} and ~\ref{fig:all:SED}, respectively. The average photon and energy flux are given by $(5.16 \pm 0.05) \times 10^{-8} \rm{cm^{-2} s^{-1}}$ and $(2.71 \pm 0.04) \times 10^{-5} \rm{MeV~cm^{-2} s^{-1}}$ respectively. 
From the SED, we find non-zero flux up to approximately 20 GeV and upper limits beyond that. Similar to segment 3A, we see an approximately constant flux, except for the first data point. We see a falling light curve from the start of segment 1 until MJD of around 58000. \rthis{We also confirmed that we do not find any signal beyond 20 GeV even aftercoarser binning.} The posteriors for the best-fit spectral parameters can be found in Fig.~\ref{fig:all:SED:par}. The isotropic average gamma-ray luminosity detected during the entire segment is $2.18 \times 10^{47}$ ergs/sec.

\rthis{We also did a search for periodicity using the data for all eight segments using the generalized Lomb-Scargle periodogram~\citep{Jake}. For this purpose, we use the same implementation as our most recent work on the analysis of Super-Kamiokande solar neutrino data~\citep{PasumariSK}. We searched using a frequency grid with a resolution of 0.028 $yr^{-1}$ and ranging from 0 to 2.69 $yr^{-1}$. The largest LS power is at 2.15 $yr^{-1}$. However, the $p$-value using all the metrics specified in ~\citet{Jake} is greater than 0.1. Therefore, we conclude that there is no discernible periodicity in the OJ 287 gamma-ray dataset during the above period.}

\begin{table}[H]
\centering
\begin{tabular}{|l|c|c|c|c|c|}
\toprule
\textbf{Segment} & \textbf{TS} & \textbf{Flux} $\times 10^{-8}$& \textbf{Energy Flux} $\times 10^{-5}$ & \textbf{Luminosity} \\
 && (cm$^{-2}$ s$^{-1}$) & (MeV cm$^{-2}$ s$^{-1}$) & $\times 10^{47}$ (erg s$^{-1}$)\\
\midrule

1 & 1586.61 & $10.79 \pm 0.70$ & $4.65 \pm 0.27$ & 3.78 \\
2 & 891.89 & $4.07 \pm 0.05$ & $3.52 \pm 0.09$ & 2.76 \\
3A & 155.29 & $3.44 \pm 0.95$ & $1.53 \pm 0.26$ & 2.76 \\
3B & 154.6 & $3.88 \pm 0.67$ & $1.86 \pm 0.33$ & 2.31 \\
4 & 263.04 & $4.80 \pm 0.78$ & $2.02 \pm 0.34$ & 0.98 \\
5 & 379.65 & $4.52 \pm 0.72$ & $2.60 \pm 0.45$ & 3.1 \\
6 & 308.48 & $3.72 \pm 0.67$ & $2.06 \pm 0.33$ & 2.18 \\
7 & 320.09 & $2.06 \pm 0.25$ & $1.95 \pm 0.27$ & 3.96 \\
8 & 173.44 & $4.61 \pm 0.18$ & $2.15 \pm 0.14$ & 1.79 \\
all & 4995.36 & $5.16 \pm 0.05$ & $2.71 \pm 0.04$ & 2.18 \\

\bottomrule
\end{tabular}
 \caption{Summary of test statistics (TS), observed photon, energy flux, and the isotropic gamma-ray luminosity. We note that the detection significance is given by the square root of TS~\citep{Mattox}.}
 \label{tab:s2:pars}
\end{table}

\begin{table}[H]
 \centering
 \begin{tabular}{|l|c|c|c||c|c|c|c|}
 \toprule
 &\multicolumn{2}{c}{\textbf{Simple PL fit}} && \multicolumn{3}{c}{\textbf{Log Parabola fit}}&\\
 \midrule
 \textbf{Segment} & \textbf{Pivot} $E_0$ (MeV) & \textbf{$\log_{10}$N$_0$} & $\alpha$ & \textbf{$\log_{10}$N$_0$} & $\alpha$ & $\beta$ & $E_b$ (MeV)\\
 \midrule
1 & $-9.51 \pm 0.03$ & $2.14 \pm 0.04$ & 636.6 & $ -10.65 \pm 0.04 $ & $ 2.15 \pm 0.04 $ & $ 0.10 \pm 0.02 $ & 710.99 \\
2 & $-9.94 \pm 0.05$ & $1.92 \pm 0.05$ & 720.59 & $ -10.97 \pm 0.01 $ & $ 1.80 \pm 0.01 $ & $ 0.07 \pm 0.00 $ & 710.99 \\
3A & $-9.89 \pm 0.16$ & $2.19 \pm 0.19$ & 846.77 & $ -11.10 \pm 0.13 $ & $ 2.11 \pm 0.16 $ & $ 0.14 \pm 0.08 $ & 710.99 \\
3B & $-10.06 \pm 0.12$ & $1.97 \pm 0.12$ & 803.72 & $ -11.16 \pm 0.10 $ & $ 2.15 \pm 0.09 $ & $ 0.03 \pm 0.04 $ & 710.99 \\
4 & $-9.91 \pm 0.06$ & $2.21 \pm 0.10$ & 545.92 & $ -11.15 \pm 0.12 $ & $ 2.27 \pm 0.09 $ & $ 0.00 \pm 0.04 $ & 710.99 \\
5 & $-9.96 \pm 0.05$ & $2.10 \pm 0.07$ & 921.45 & $ -11.11 \pm 0.09 $ & $ 2.13 \pm 0.08 $ & $ 0.00 \pm 0.03 $ & 710.99 \\
6 & $-9.97 \pm 0.06$ & $2.03 \pm 0.08$ & 745.69 & $ -11.11 \pm 0.11 $ & $ 2.04 \pm 0.10 $ & $ 0.05 \pm 0.05 $ & 710.99 \\
7 & $-10.20 \pm 0.13$ & $1.84 \pm 0.12$ & 1081.02 & $ -11.19 \pm 0.09 $ & $ 1.69 \pm 0.07 $ & $ 0.12 \pm 0.03 $ & 710.99 \\
8 & $-10.07 \pm 0.07$ & $1.96 \pm 0.08$ & 705.82 & $ -11.26 \pm 0.03 $ & $ 2.33 \pm 0.02 $ & $ -0.07 \pm 0.01 $ & 710.99 \\
all & $-9.87 \pm 0.01$ & $2.12 \pm 0.02$ & 715.56 & $ -11.02 \pm 0.01 $ & $ 2.11 \pm 0.01 $ & $ 0.03 \pm 0.00 $ & 710.99 \\

 \bottomrule
 \end{tabular}
 \caption{\rthis{Results of spectral fits using the power-law (cf. Eq.~\ref{eq:1}) as well as log-parabola model (cf. Eq.~\ref{eq:logparabola}) for all the eight segments}}
 \label{tab:my_label}
\end{table}

\section{Comparison with other monitoring programs for OJ 287}
\label{sec:aux}
\subsection{Comparison with Prince et al.}
We now compare our results with the gamma-ray observations for OJ 287 reported in ~\citet{Prince}. A summary of their observations can be found in Table 4. We find that the spectral index $\alpha$ is in the same ball-park range between 1.85 and 2.24 found in ~\citet{Prince}. The photon flux in segment 1 is about a factor of two larger than the fluxes detected in \citet{Prince}. For all other segments, it is comparable. The isotropic luminosity that we detect, which is between $10^{47}-10^{48}$ ergs/sec, is comparable to that detected in ~\cite{Prince} and is less than the Eddington luminosity of $10^{50}$ erg/sec. Therefore, our results are mostly consistent with those in ~\citet{Prince}.

\begin{figure}
 \centering
 \includegraphics[width=0.7\textwidth]{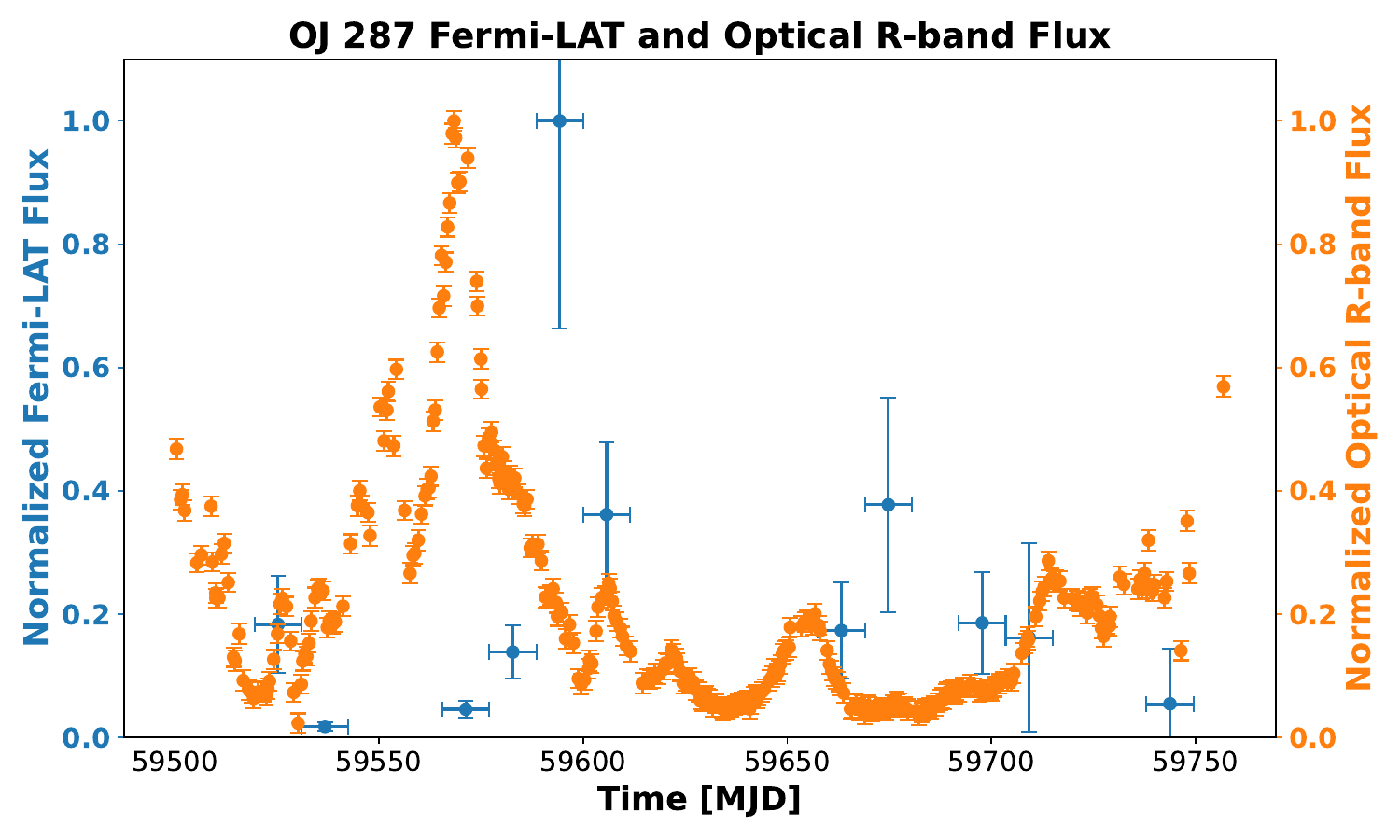}
 \caption{\rthis{Comparison of the optical $R$-band observations reported in ~\citet{Valtonen24} alongside the flux observed by Fermi-LAT during Segment 7. The gamma-ray and optical light curves have been normalized by their respective modes. }}
 \label{fig:R_LC_SUP}
\end{figure}

\subsection{Comparison with $R$-band optical flare in 2021-2022}
\label{sec:opticalgammacomp}
Most recently, ~\citet{Valtonen24} reported optical flares in the $R$-band between October 21st and November 2021 as part of the Krakow quasar monitoring campaign consisting of several telescopes located all over the globe. These flares are caused by the impact of the secondary black hole onto the accretion disk of the primary black hole. This flare occurred during segment 7. To compare the optical data with the Fermi-LAT data, we plot both in the same figure after normalizing each data by the mode. This plot can be found in Fig.~\ref{fig:R_LC_SUP}. As we can see, although the gamma-ray light curve is sampled coarsely compared to the optical light curve, there is no direct one-to-one correlation between the rise and fall time of the gamma-ray and optical light curves during this segment. The fall in the optical light curve coincides with the rising gamma-ray flux at the beginning of the segment, where the peaks in the two light curves are separated by around 200 days. \vfill
\subsection{Comparison with SWIFT XRT light curves from 2015-2021}
\begin{figure}[h]
 \centering
 \includegraphics[width=0.7\textwidth]{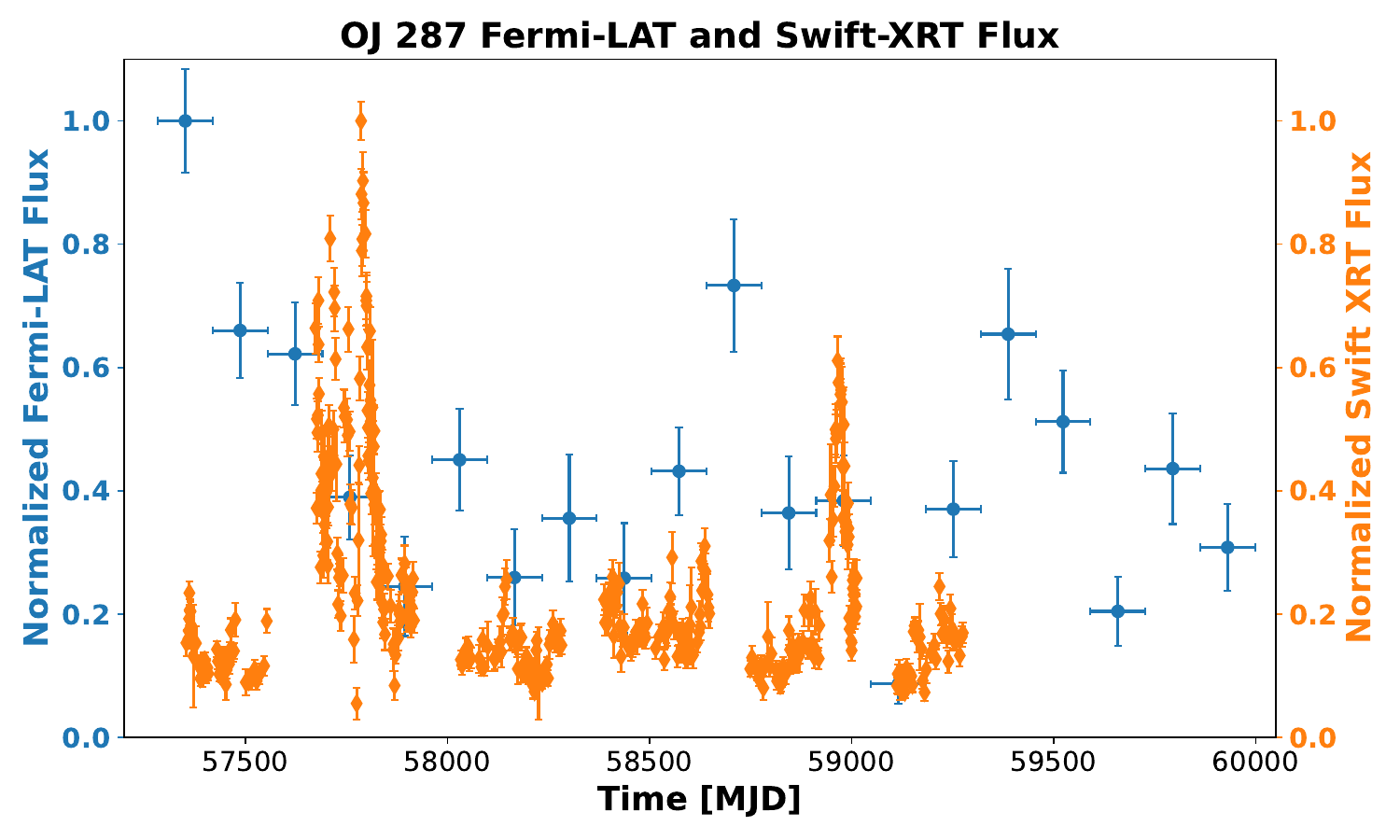}
 \caption{Comparison of the SWIFT XRT observations from 0.3-10 keV obtained from~\citet{Komossa22}, alongside the flux observed by Fermi-LAT. The gamma-ray and X-ray observations are normalized by their respective modes.}
 \label{fig:XRT_LC_SUP}
\end{figure}

OJ 287 has also been monitored by the SWIFT satellite in optical/UV bands using six filters as well as X-rays from 0.3-10 keV as part of the MOMO (Multiwavelength Observations and Modelling of OJ 287) project~\citep{MOMO} from September 2015 till September 2021. Additional SWIFT observations starting from 2015 have also been reported~\citep{Komossa20,Komossa21}.
These observations have been collated in ~\citet{Komossa22}. Theoretical implications of the observations from the MOMO project can be found in ~\citet{Komossa23,MOMO6}.
We overlay these SWIFT X-ray observations along with the Fermi-LAT observations in Fig.~\ref{fig:XRT_LC_SUP}. Once again we do not see any correlation between the peaks in the SWIFT XRT light curves and gamma-ray light curves.

\section{Discussion and Conclusions}
\label{sec:conclusions}
In a recent work,~\citet{Gupta23} presented the results of an extensive observing campaign of OJ 287 from 2015-2023 using optical telescopes in three different continents in conjunction with public archival data from the Steward Observatory. These observations were divided into eight segments, each roughly corresponding to an observing season. The third segment was further subdivided into two sub-segments. We searched for coincident gamma-ray emission from 0.1-300 GeV with the Fermi-LAT telescope during each observing segment. For each segment, we show 
the gamma-ray light curve and SED. We fit the SED to a power law and show a corner plot depicting the best-fit parameters. We also calculate the luminosity for each segment. the plots for the gamma-ray light curves, SED, and marginalized contours for the SED best-fit parameters for all the eight segments can be found in Figs.~1-30. Tables ~\ref{tab:s2:pars} and ~\ref{tab:my_label} provides a tabular summary of our results. Our conclusions are as follows:
\begin{itemize}
 \item We detect non-zero gamma-ray emission ($> 10\sigma$) during all the eight segments, with flux between $(2-10) \times 10^{-8} \rm{cm^{-2} s^{-1}}$
 \item The observed energy flux is between $(2-4.6) \times 10^{-5}~\rm{MeV cm^{-2} s^{-1}}$.
 \item Ths isotropic luminosity is between $(0.98-3.96) \times 10^{47}$ ergs/sec.
 \item The power-law index varies between 1.8 - 2.2.
 \item The maximum flux is seen near the start of the first segment. The flux decreases until MJD of 58000. 
 \item We do not find any correlated gamma-ray flare during Segment 7 in coincidence with an optical flare in $I$ and $R$ band recently reported in ~\citet{Valtonen24}. A comparison of the gamma-ray and optical light curves can be found in Fig.~\ref{fig:R_LC_SUP}.
 \item We also do not find any correlation between the peaks in the gamma-ray flux and X-ray flux observed by the SWIFT satellite between 0.3-10 keV (cf. Fig~\ref{fig:XRT_LC_SUP}).

 \item We do not detect any emission beyond 20-30 GeV, similar to previously reported gamma-ray observation from OJ287~\citep{Prince,Kushwaha13}. The gamma-ray spectrum in OJ287 has been previously explained due to a combination of Synchrotron self-Compton and external Compton processes from the scattering of soft photons that are outside the jet~\citep{Kushwaha13}. The cut-off in the spectrum can be explained with a power law distribution of electrons with a maximum energy of approximately $10^4$ times the electron rest mass~\cite{Kushwaha13}. 
\end{itemize}
\section*{Acknowledgments}
We are grateful to Raniere Menezes for his help with the EasyFermi package. We are grateful to Stefanie Komossa for providing us the SWIFT X-ray data from ~\citet{Komossa22} and Mauri Valotnen for $R$-band data reported in ~\citet{Valtonen24}. We also acknowledge the anonymous referee for constructive feedback on our manuscript.
\bibliography{main.bib}
\end{document}